\documentclass[prl,showpacs,twocolumn,superscriptaddress]{revtex4}
\usepackage{graphicx}
\usepackage{psfig}
\usepackage{epsfig}
\usepackage{bm}
\usepackage{amsmath}
\newcommand{\eep}{(e,e$'$p)}
\newcommand{\bra}[1]{\left\langle #1\right|}
\newcommand{\ket}[1]{\left| #1\right\rangle}
\begin{document}

\title{Saturation of nuclear matter and short-range correlations}

\author{Y. Dewulf}
\affiliation{Laboratory of Theoretical Physics, Ghent University, 
Proeftuinstraat 86, B-9000 Gent, Belgium}

\author{W. H. Dickhoff}
\affiliation{Department of Physics, Washington University, St. Louis, 
Missouri 63130, USA}
\affiliation{Laboratory of Theoretical Physics, Ghent University, 
Proeftuinstraat 86, B-9000 Gent, Belgium}

\author{D. Van Neck}
\affiliation{Laboratory of Theoretical Physics, Ghent University, 
Proeftuinstraat 86, B-9000 Gent, Belgium}

\author{E. R. Stoddard}
\affiliation{Department of Physics, University of Missouri at
Kansas City, Kansas City, Missouri 64110, USA}

\author{M. Waroquier}
\affiliation{Laboratory of Theoretical Physics, Ghent University, 
Proeftuinstraat 86, B-9000 Gent, Belgium}

\date{\today}

\begin{abstract}
A fully self-consistent treatment of short-range correlations in
nuclear matter is presented.
Different implementations of the determination of the nucleon spectral
functions for different interactions are shown to be consistent with each 
other.
The resulting saturation densities are closer to the empirical
result when compared with (continuous choice) Brueckner-Hartree-Fock values. 
Arguments for the dominance of short-range correlations 
in determining the nuclear-matter saturation density are presented.
A further survey of the role of long-range correlations suggests that the
inclusion of pionic contributions to ring diagrams in nuclear matter leads to
higher saturation densities than empirically observed.
A possible resolution of the nuclear-matter saturation problem
is suggested.
\end{abstract}

\pacs{21.65.+f, 21.30.Fe}

\maketitle

A correct description of the saturation properties of nuclear matter
has remained an unresolved issue for a very long time.
The Brueckner-Bethe-Goldstone (BBG) expansion~\cite{day1} supplies a converged
result for the energy per particle in the relevant density range,
for a given realistic interaction, at the level of three hole-line
contributions~\cite{day2,bald}.
Such calculations fail to reproduce the empirical saturation properties
which require a minimum in the equation of state at a density corresponding
to a Fermi momentum, $k_F$, of about 1.33 fm$^{-1}$ with a binding energy
of about 16 MeV.
The authors of Ref.~\cite{bald} obtain for the Argonne $v_{14}$ 
interaction~\cite{wsa84} a saturation density corresponding
to 1.565 fm$^{-1}$ with about the correct amount of binding.
This corresponds to an overestimation of the empirical density by about
60\% but appears completely consistent with corresponding variational 
calculations~\cite{dayw} for the same interaction.

Several different remedies for this serious problem have been proposed
over the years. The intrinsic structure of the nucleon and its related
strong coupling to the $\Delta$-isobar inevitably requires the consideration
of three-body (or more-body) forces. When three-body forces are considered
in variational calculations it is possible to achieve better saturation
properties only when an adhoc repulsive short-range component of this
three-body force is added~\cite{cpw,wff}.
It has also been suggested that a relativistic treatment of the nucleon
in the medium using a Dirac-Brueckner approach provides the necessary
ingredients for a better description of saturation~\cite{acps,hama,brma,amtj}. 

All many-body methods developed for nuclear matter have focused on a
proper treatment of short-range correlations (SRC) without
the benefit of experimental information on the influence of these
correlations on the properties of the nucleon in the medium.
This influence can now be clearly identified by considering recent results
from \eep\ reactions~\cite{diehu,sihu,lap1} and theoretical calculations
of the nucleon spectral function in nuclear matter~\cite{rpd,bff,vond1}.
A recent analysis of the \eep\ reaction on ${}^{208}$Pb
in a wide range of missing energies and for missing momenta below 270 MeV/c
yields information on the occupation numbers of all the
deeply-bound proton orbitals.
These data indicate that all these orbitals are depleted by the same
amount of about 15\%~\cite{lap2}.
These occupation numbers are associated with
the orbits which yield an accurate fit to the \eep\ cross section.
The properties of these occupation numbers suggest that the main effect of
the global depletion of these mean-field orbitals is due to SRC.
Indeed, the effect of the coupling
of hole states to low-lying collective excitations only affects occupation
numbers of states in the immediate vicinity of the Fermi energy~\cite{rijs}.
In addition, nuclear matter momentum distributions display such
an overall global depletion due to short-range and tensor
correlations~\cite{vond1,vond2,fapa}.
The latter results formed the basis of the now corroborated 
prediction~\cite{dimu1,papa}
for the occupation numbers in ${}^{208}$Pb~\cite{lap2}.

Most of this depleted single-particle (sp) strength is located at 
energies more than 100 MeV
above the Fermi energy~\cite{vond1,vond2,dimu1}.
This appearance of strength at high energy is another important aspect of
the influence of short-range and tensor correlations.
Yet another characteristic feature of these SRC is that
this depletion of the sp strength
must be compensated by the admixture of a corresponding number
of particles with high-momentum components.
These high-momentum components have not yet been unambiguously identified
but are currently studied experimentally~\cite{rohe}.
Solid theoretical arguments~\cite{ciofi} and calculations clearly
pinpoint this strength at high excitation energy in the hole 
spectrum both for nuclear matter~\cite{vond1} and finite
nuclei~\cite{mpd1}.
Indeed, experiment confirms that no substantial admixture of these high-momentum
components is observed in the vicinity of the Fermi energy~\cite{bobel}.

We now present an argument showing that SRC are the dominant
factor in determining the empirical saturation density of nuclear matter.
We recall that elastic electron scattering from
${}^{208}$Pb~\cite{froi} accurately determines
the value of the central charge density in this nucleus.
By multiplying this number by $A/Z$ one obtains the relevant central density
of heavy nuclei, corresponding to 0.16 nucleons/fm${}^3$ or $k_F = 1.33$
fm${}^{-1}$.
Since the presence of nucleons at the center of a heavy nucleus is confined
to $s$-wave nucleons, and, as discussed above, their depletion is
dominated by SRC, one may therefore conclude that the same is true for 
the actual value of the empirical saturation density of nuclear matter.
While this argument is particularly appropriate for the deeply bound
$1s_{1/2}$ and $2s_{1/2}$ protons, it continues to hold to
a large extent for the $3s_{1/2}$
protons which are depleted predominantly by short-range effects (up to 15\%)
and by at most 10\% due to long-range correlations~\cite{sihu,grab}.
These considerations demonstrate clearly that one may expect
SRC to have a decisive influence on the actual value
of the nuclear-matter saturation density.

High-momentum components due to SRC also have a 
considerable impact on the binding energy of nuclear matter.
This result can be inferred from the energy sum
rule~\cite{gami}
\begin{equation}
\frac{E}{A} = \frac{2}{\rho} \int \frac{d^3k}{(2\pi)^3}
\int_{-\infty}^{\varepsilon_F} d\omega\ \left( \frac{k^2}{2m} +
\omega \right) S_h(k,\omega) ,
\label{eq:epp}
\end{equation}
where $\rho = \frac{2k_F^3}{3\pi^2}$ is the density.
Eq.~(\ref{eq:epp}) illustrates the link between the energy of the system and
the hole spectral function, $S_h(k,\omega)$.
Results for the momentum distribution and true potential energy
based on the spectral function show that enhancements
as large as 200\% for the kinetic and 
potential energy over the mean-field values can be obtained
for both nuclear matter~\cite{vond2} and finite nuclei~\cite{mpd1}.
These large attractive contributions to the potential energy
of nuclear matter are mainly from weighing the 
high-momentum components in the spectral function with 
large negative energies in Eq.~(\ref{eq:epp}).
The location of these high-momentum components as a function of energy
is therefore an important ingredient in the determination of the energy
per particle as a function of density.
So far, the determination of this location has relied only on
quasiparticle properties in the construction of the self-energy.
A self-consistent determination of the spectral function
including the location of these high-momentum 
components therefore includes the dominant physics of SRC in 
the description of nuclear matter and is consistent with the
experimental observations of the nucleon spectral function in nuclei.

Such a determination
requires the solution of the ladder equation for the effective interaction
in the medium
\begin{eqnarray}
\label{eq:ladpw}
\bra{q}\Gamma_{\ell\ell'}^{JST}(K,\Omega)\ket{q'}
& = & \bra{q}V_{\ell\ell'}^{JST}\ket{q'} \\
& + & \sum_{\ell''} \int_0^\infty dpp^2
\bra{q}V_{\ell\ell''}^{JST}\ket{p} \nonumber \\
& \times & \bar{\mathsf{g}}^{II}_f(p;K,\Omega)
\bra{p}\Gamma_{\ell''\ell'}^{JST}(K,\Omega)\ket{q'} , \nonumber
\end{eqnarray}
where a notation with relative momenta $p,q,q'$ and the conserved total 
momentum $K$
has been used. The propagator $\bar{\mathsf{g}}^{II}_f$ in 
Eq.~(\ref{eq:ladpw}) has been obtained by an angle-averaging procedure of the
noninteracting two-particle propagator
\begin{eqnarray}
\mathsf{g}^{II}_f(k,k';\Omega) & =  &
\int_{\varepsilon_F}^{\infty} \!\! d\omega \int_{\varepsilon_F}^\infty
\!\! d\omega' \frac{ S_p(k, \omega)
S_p(k', \omega') }
{\Omega -\omega -\omega' +i\eta} \nonumber \\ 
& - &
\int_{-\infty}^{\varepsilon_F} \!\!d\omega \int_{-\infty}^{\varepsilon_F}
\!\!d\omega' \frac{ S_h(k, \omega)
S_h(k', \omega') }
{\Omega -\omega -\omega' -i\eta} , 
\label{eq:g2fdis}
\end{eqnarray}
in order to allow a partial wave decomposition of the ladder equation.
Note that sp momenta $k,k'$ are used in Eq.~(\ref{eq:g2fdis}).
In turn, the spectral functions are needed to determine Eq.~(\ref{eq:g2fdis})
and thereby the effective interaction $\Gamma$ through Eq.~(\ref{eq:ladpw}).
They can be obtained from the imaginary part of the sp propagator which
solves the Dyson equation
\begin{equation}
\mathsf{g}(k,\omega) = \mathsf{g}^{(0)}(k,\omega)
+ \mathsf{g}^{(0)}(k,\omega) \Sigma(k,\omega) \mathsf{g}(k,\omega) ,
\label{eq:dys}
\end{equation}
where the self-energy $\Sigma$ includes the contribution of SRC through 
$\Gamma$, to complete the self-consistency loop.

The implementation of this self-consistency scheme is numerically
quite involved and
has been attempted by several groups~\cite{dejong,rd,dwvnw,bozek0,
bozek1,bozek2}.
In the present paper two different approaches have been used to generate
results for different interactions.
In the continuous scheme, 
a representation of the imaginary part of the self-energy
in terms of four gaussians is used to completely describe the sp
propagator. The parameters of these gaussians are then determined
self-consistently ~\cite{rd} for the Reid potential~\cite{reid} by solving
Eq.~(\ref{eq:ladpw}) with the convolution of spectral functions
in Eq.~(\ref{eq:g2fdis}) as input, constructing the self-energy,
and then solving the Dyson equation (\ref{eq:dys}).
In the discrete scheme we used a representation of the propagator
in terms of three discrete poles \cite{dwvnw}, which avoids a full continuum 
solution of Eq.~(\ref{eq:ladpw}).
The latter approach is equivalent to a continuous version as far as the 
energy per particle is concerned, since it 
requires  a reproduction of the relevant energy-weighted moments of the 
hole and particle spectral function~\cite{dwvnw}.
This is substantiated by comparing the results of this discrete scheme with 
the results of the continuous self-consistency scheme used in 
Ref.~\cite{bozek0} for the Mongan-type separable interaction \cite{mon} 
and in Ref.\cite{bozek2} for the separable Paris interaction \cite{haid}.
We find that the binding energies correspond to 
within 5\%  over the relevant $k_F$ range around the minimum, and moreover 
that the location of the minimum agrees to within 3\%.

\begin{figure}
\vspace*{-1.0cm}       
\includegraphics[width=0.5\textwidth,height=0.5\textwidth] {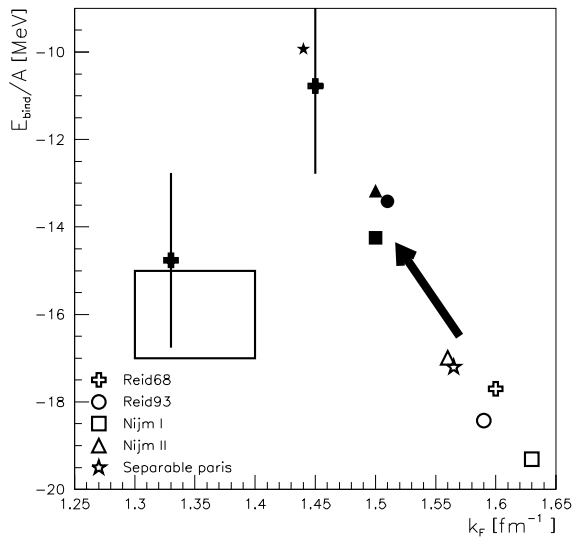}
\vspace*{-.8cm}       
\caption{\label{fig:coester} Nuclear matter
saturation points calculated with various realistic
NN interactions. The open symbols refer to continuous choice 
Brueckner-Hartree-Fock results. The filled symbols refer to
self-consistent results and represent saturation points calculated in the 
discrete scheme, except for the Reid68
interaction where the binding energy at two densities is shown in the 
continuous scheme. 
 }
\end{figure}

In Fig.\ref{fig:coester} we report the saturation points obtained within the 
discrete scheme of Ref.~\cite{dwvnw}
for the updated Reid potential (Reid93), the NijmI
and NijmII interaction ~\cite{reid93} and the separable Paris interaction
\cite{haid}. 
The results demonstrate 
an important and systematic change of the saturation properties with respect 
to continuous choice Brueckner-Hartree-Fock (ccBHF) calculations, 
leading to about 4-6 MeV less binding, 
and reduced values of the saturation density, closer to the empirical one.   
Such a trend is entirely consistent with the observations in 
Ref.~\cite{bozek2} made for a separable NN interaction, and is now extended 
to more realistic (non-separable) interactions.   

The discrete scheme~\cite{dwvnw} could not be used with the original 
Reid (Reid68) potential because of its slow decay in momentum space,  
but some results are available in the continuous scheme of  
Ref.~\cite{rd}. In Fig.\ref{fig:coester} the binding energy is shown 
at two densities ($k_F = 1.33$ and 1.45 fm$^{-1}$); 
the error bars are an estimate of the remaining uncertainty due to incomplete 
convergence and the non-selfconsistent treatment of some higher
order partial waves \cite{rd}. 
The results again seem to indicate a substantial shift in the 
saturation density for the Reid68 potential, from the ccBHF value of about 
1.6 fm$^{-1}$, to a value below 1.45 fm$^{-1}$, without seriously 
underbinding nuclear matter.

The present self-consistent treatment of SRC (scSRC) differs in two main 
aspects from the ccBHF approach. Firstly, hole and particle lines are treated 
on an equal footing, thereby ensuring thermodynamic consistency\cite{bozek0}. 
Intermediate hole-hole propagation in the ladder diagrams is included to all 
orders. This feature provides, compared to ccBHF, a substantial repulsive 
effect in the $k<k_F$ contribution to Eq.~(\ref{eq:epp}), and 
 comes primarily from an upward shift of the quasi-particle 
energy spectrum as a result of including $\omega < \epsilon_F$ 
contributions to the imaginary part of the self-energy.
The effect increases with density, and is the dominant factor in the 
observed shift of the saturation point. 
Secondly, the realistic spectral functions, generated through 
Eqs.\mbox{(\ref{eq:ladpw}-\ref{eq:dys})} and used in the evaluation of the
in-medium interaction $\Gamma$ and self-energy $\Sigma$, are in agreement 
with experimental information obtained from \eep\ reactions.  
For the Reid93 interaction at $k_F=1.37$ fm$^{-1}$ we find $z=0.74$ for 
the quasiparticle strength at the Fermi momentum, whereas the hole 
strength for $p=0$, integrated up to 100 MeV missing energy, equals 83\%;
similar values are found for the other interactions.  
The depletion of the quasiparticle peaks is primarily important to 
suppress unrealistically large pairing instabilities around normal density.
The improved treatment of the high-momentum components does 
affect the binding energy, through the $k>k_F$ contribution to Eq.~(\ref{eq:epp}).  
This feature, studied in \cite{dwvnw}, provides a sizeable 
attraction, but is smaller than the afore-mentioned repulsive effect.

The inclusion of $hh$-propagation in scSRC also leads 
to a somewhat stiffer equation-of-state than in ccBHF. A recent analysis of 
the giant monopole resonance in heavy nuclei 
\cite{blaizot} yields an experimental estimate  
$K_{nm}=210\pm 30$MeV for the 
nuclear matter compression modulus,
\begin{equation}
K_{nm}=\left. k_F^2\frac{d^2 E/A}{dk_F^2} \right|_{k_F=k_{F,0}}.
\end{equation}
At the saturation points in Fig.\ref{fig:coester} we find ccBHF values  
$K_{nm}=154$ MeV for Reid93 and $K_{nm}=148$ MeV for the 
separable Paris interaction, which are enhanced to $K_{nm}=177$ MeV 
and $K_{nm}=216$ MeV, respectively, in our scSRC calculation.  
These values agree reasonably well with the experimental estimate. 
Note that reasonable values for $K$ imply that the Reid68 energies in 
Fig.\ref{fig:coester} 
may still deviate by 1-1.5 MeV from numerically exact scSRC 
values, as indicated by the error bars.

The present results indicate that a sophisticated treatment 
of SRC lowers the ccBHF saturation densities, bringing them closer to the 
empirical one. 
It remains to be understood why apparently converged hole-line 
calculations~\cite{bald} yield higher saturation densities.
The three hole-line terms obtained in Ref.~\cite{bald} 
indicate reasonable convergence properties compared to
the two hole-line contribution. One may therefore assume that
these results provide an accurate representation of the energy per particle
of nuclear matter as a function of density for the case of nonrelativistic
nucleons and two-body forces.
At this point it is useful to identify an underlying assumption when the 
nuclear-matter problem is posed \cite{rd}.
This assumption asserts that the influence of long-range correlations
in finite nuclei and nuclear matter are commensurate. 
We'd like to point out that this underlying assumption is questionable. 
Three
hole-line contributions include a third-order ring diagram characteristic
of long-range correlations.
The effect of long-range correlations on nuclear saturation properties
is sizeable, as shown by the results for three-
and four-body ring diagrams calculated in Ref.~\cite{dfm}
(see also Refs.~\cite{day2,bald}).
The results of Ref.~\cite{dfm} demonstrate that such ring-diagram terms
are dominated by attractive contributions involving pion quantum numbers 
propagating around the rings, and increase in importance with increasing 
density. Such long-range pion-exchange contributions to the
binding energy appear due to the possibility to coherently
sample the attractive interaction in a given ring diagram at 
momenta $q$ above 0.7 fm$^{-1}$.
This feature is related to momentum conservation in nuclear
matter and is not available in finite nuclei, in which 
no such collective pion-degrees of freedom are actually observed~\cite{cz2}.
It seems therefore reasonable to call into question the relevance
of these coherent long-range pion-exchange contributions
to the binding energy per particle since their behavior is so markedly
different in finite and infinite systems.
One may consider the salient difference of the ratio of spin-longitudinal
and spin-transverse response function in nuclear matter and finite
nuclei as another indication of the relevance of our suggestion~\cite{alber}.
We also like to point out that experimental information of these response 
functions~\cite{carey,tadd,wakas}
suggests no characteristic enhancement of the (pionic) spin-longitudinal
response as expected on the basis of nuclear matter calculations.

Clearly, the assertion that long-range pion-exchange contributions
to the energy per particle need not be considered in explaining nuclear
saturation properties, needs to be further investigated.
At this point  it appears that a fully self-consistent treatment of SRC has 
substantially different saturation properties than a conventional 
(continuous choice) Brueckner-Hartree-Fock treatment, and is capable of 
yielding saturation densities close to the empirical one.  

\vspace{.5truecm}
This work is supported by the U.S. National Science Foundation under Grant
Nos. PHY-9900713 and PHY-0140316 and by the Fund for Scientific 
Research-Flanders (FWO-Vlaanderen) and the Research Council of Ghent 
University.


\begin{thebibliography}{99}

\bibitem{day1} B. D. Day, Rev. Mod. Phys. \textbf{39}, 719 (1967).

\bibitem{day2} B. D. Day, Phys. Rev. C \textbf{24}, 1203 (1981).

\bibitem{bald} H. Q. Song, M. Baldo, G. Giansiracusa, and U. Lombardo,
Phys. Rev. Lett. \textbf{81}, 1584 (1998).

\bibitem{wsa84} R. B. Wiringa, R. A. Smith, and T. L. Ainsworth,
Phys. Rev. C \textbf{29}, 1207 (1984).

\bibitem{dayw}
B. D. Day and R. B. Wiringa, Phys. Rev. C \textbf{32}, 1057 (1985).

\bibitem{cpw} J. Carlson, V. R. Pandharipande, and R. B. Wiringa, Nucl. Phys.
\textbf{401}, 59 (1983).

\bibitem{wff} R. B. Wiringa, V. Fiks, and A. Fabrocini, Phys Rev. C
\textbf{38}, 1010 (1988).

\bibitem{acps} M. R. Anastasio, L.S. Celenza, W. S. Pong, and C. M. Shakin,
Phys. Rep. \textbf{100}, 327 (1983).

\bibitem{hama} B. ter Haar and R. Malfliet, Phys. Rev. Lett. \textbf{56},
1237 (1986).

\bibitem{brma} R. Brockmann and R. Machleidt, Phys. Rev. C \textbf{42},
1965 (1990).

\bibitem{amtj} A. Amorim and J. A. Tjon, Phys. Rev. Lett. \textbf{68}, 772
(1992).

\bibitem{diehu} A. E. L. Dieperink and P. K. A. deWitt Huberts,
Ann. Rev. Nucl. Part. Sci. \textbf{40}, 239 (1990).

\bibitem{sihu} I. Sick and P. K. A. deWitt~Huberts,
Comm. Nucl. Part. Phys. \textbf{20}, 177 (1991).

\bibitem{lap1} L. Lapik{\'{a}}s, Nucl. Phys. \textbf{A553}, 297c (1993).

%\bibitem{pasi} V.R. Pandharipande, I. Sick, and P.K.A. deWitt Huberts, 
%Rev. Mod. Phys. \textbf{69}, 981 (1997)

\bibitem{rpd} A. Ramos, A. Polls, and W. H. Dickhoff, Nucl. Phys.
\textbf{A503}, 1 (1989).

\bibitem{bff} O. Benhar, A. Fabrocini, and S. Fantoni, Nucl. Phys.
\textbf{A505}, 267 (1989).

\bibitem{vond1} B.E. Vonderfecht, W.H. Dickhoff, A. Polls, and
A. Ramos, Phys. Rev. C \textbf{44}, R1265 (1991).

\bibitem{lap2} L. Lapik{\'{a}}s \textit{et al.}, to be submitted to Phys. Rev.
Lett. (2002);
M. F. van Batenburg, Ph.D. thesis, University of Utrecht (2001).

\bibitem{rijs} G.~A.~Rijsdijk, K.~Allaart, and W.~H.~Dickhoff,
Nucl. Phys. \textbf{A550}, 159 (1992).

\bibitem{vond2} B. E. Vonderfecht, W. H. Dickhoff, A. Polls, and
A. Ramos, Nucl. Phys.\textbf{A555}, 1 (1993).

\bibitem{fapa} S. Fantoni and V. R. Pandharipande, Nucl. Phys. \textbf{A427},
473 (1984).

\bibitem{dimu1} W. H. Dickhoff and H. M{\"{u}}ther, Rep. Prog. Phys.
\textbf{55}, 1947 (1992).

\bibitem{papa} V.R. Pandharipande, C.N. Papanicolas, and J. Wambach, 
Phys. Rev. Lett. \textbf{53}, 1133 (1984)
 
\bibitem{rohe} D. Rohe, Proc. 5$^{th}$ Workshop on ``e-m induced
Two-Hadron Emission'', 2001, Lund, Sweden (CD ISBN:91-631-1612-X) 67.
%, http://2nconf.nuclear.lu.se; 
%D. Rohe {\em et al.}, to be published (2002). 

\bibitem{ciofi} C. Ciofi degli Atti, S. Liuti, and S. Simula,
Phys. Rev. C \textbf{41}, R2474 (1990).

\bibitem{mpd1} H. M{\"{u}}ther, A. Polls, and W. H. Dickhoff, Phys. Rev.
C \textbf{51}, 3040 (1995).

\bibitem{bobel}I. Bobeldijk \textit{et al.}, Phys. Rev. Lett.
\textbf{73}, 2684 (1994).

\bibitem{froi} B.~Frois \textit{et al.}, Phys. Rev. lett \textbf{38},
152 (1977).

\bibitem{grab} P. Grabmayr, Prog. Part. Nucl. Phys. \textbf{29}, 251 (1992).


\bibitem{gami} V. M. Galitski and A. B. Migdal, Sov. Phys. JETP 
\textbf{34}, 96 (1958).

%\bibitem{kolt1} D. S. Koltun, Phys. Rev. Lett. \textbf{28}, 182 (1972);
%Phys. Rev. C \textbf{9}, 484 (1974).

\bibitem{dejong}
F. de Jong and H. Lenske, Phys. Rev.{\bf C56} (1997) 154

\bibitem{rd} W. H. Dickhoff and E. P. Roth, Acta Phys. Pol. B
\textbf{33}, 65 (2002); E. P. Roth, Ph.D. thesis Washington University,
St. Louis (2000).

\bibitem{dwvnw} Y. Dewulf, D. Van Neck, and M. Waroquier, Phys. Rev.
C \textbf{65}, 054316 (2002).

\bibitem{bozek0} P. Bo{\.{z}}ek and P. Czerski, Eur. Phys. J. A \textbf{11},
271 (2001).

\bibitem{bozek1} P. Bo{\.{z}}ek, Phys. Rev. C \textbf{65}, 054306 (2002).
\bibitem{bozek2} P. Bo{\.{z}}ek, Eur. Phys. J. A \textbf{15}, 325 (2002).
\bibitem{reid} R. V. Reid, Ann. Phys. \textbf{50}, 411 (1968).
\bibitem{mon} T. R. Mongan, Phys. Rev. \textbf{178}, 1597 (1969).
\bibitem{reid93} V. G. J. Stoks, R. A. M. Klomp, C. P. F. Verheggen,
and J. J. de Swart, Phys. Rev. C \textbf{49}, 2950 (1994).
\bibitem{haid} J.Haidenbauer and W.Plessas, Phys. Rev C \textbf{30}, 1822
 (1984); Phys. Rev C \textbf{32}, 1424 (1985). 
\bibitem{blaizot}  J.P.Blaizot, J.F.Berger, J.Decharg\'e and M.Girod,
Nucl.Phys. \textbf{A591}, 435 (1995)
%\bibitem{jls} A.~D.~Jackson, A.~Land{\'{e}}, and R.~A.~Smith,
%Phys. Rep. \textbf{86}, 55 (1982).

\bibitem{dfm}
W. H. Dickhoff, A. Faessler, and H. M{\"{u}}ther,
Nucl. Phys.\textbf{A389}, 492 (1982).

%\bibitem{dickdel}
%W. H. Dickhoff, Prog. Part. Nucl. Phys. \textbf{12}, 529 (1983).

\bibitem{cz2}
P. Czerski, W. H. Dickhoff, A. Faessler, and H. M{\"{u}}ther,
Phys. Rev. C \textbf{33}, 1753 (1986).

\bibitem{alber}
W. Alberico, \textit{et al},
Phys. Rev. C \textbf{34}, 977 (1986).

\bibitem{carey}
T. A. Carey \textit{et al}., Phys. Rev. Lett. \textbf{53}, 144 (1984).

\bibitem{tadd}
T. N. Taddeucci \textit{et al}., Phys. Rev. Lett. \textbf{73}, 3516 (1994).

\bibitem{wakas}
T. Wakasa \textit{et al}., Phys. Rev. C \textbf{59}, 3177 (1999).

\end{thebibliography}
\end{document}